# Dynamic Metadata Schemes in the Neutron and Photon Science Communities: A Case Study of X-Ray Photon Correlation Spectroscopy

Amir Tosson, Mohammad Reza, Christian Gutt

*Abstract*—Metadata is one of the most important aspects for advancing data management practices within all research communities. Definitions and schemes of metadata are inter alia of particular significance in the domain of neutron and photon scattering experiments covering a broad area of different scientific disciplines. The demand of describing continuously evolving highly non-standardized experiments, including the resulting processed and published data, constitutes a considerable challenge for a static definition of metadata. Here, we present the concept of dynamic metadata for the neutron and photon scientific community, which enriches a static set of defined basic metadata. We explore the idea of dynamic metadata with the help of the use case of X-ray Photon Correlation Spectroscopy (XPCS), which is a synchrotron-based scattering technique that allows the investigation of nanoscale dynamic processes. It serves here as a demonstrator of how dynamic metadata can improve data acquisition, sharing, and analysis workflows. Our approach enables researchers to tailor metadata definitions dynamically and adapt them to the evolving demands of describing data and results from a diverse set of experiments. We demonstrate that dynamic metadata standards yield advantages that enhance data reproducibility, interoperability, and the dissemination of knowledge.

## I. INTRODUCTION

IN scientific research, data is undergoing a transformative increase in both size and flow. Therefore, the quality and depth of metadata has become crucial for organizing data and sharing knowledge [1]. In the field of photon and neutron science, a set of metadata shall at least describe the basic experimental parameters such as photon energy, sample detector distance, detector etc. and the sample system investigated. However, the details of experimental setups and sample systems vary quite strongly from beamline to beamline and even within a beamline several experimental techniques are usually offered to the users. This challenge of depth, precision and volume of metadata is increased by the fact that many experiments are often non-standardized with set-ups and samples installed only for a few days and then transported back to the home laboratory after the beam time. Several user consortia started to work on this problem, such as PANSOC, EXPANDS and DAPHNE4NFDI to agree within the user community on a basic set of metadata needed [2].

A basic metadata scheme usually describes the process of raw data production. In many instances, however, the processed data and the finally published data are also of great if not even more interest to the scientific user community. In this context of data analysis and re-processing raw data, the scientific details start to matter which are difficult to describe in a static metadata scheme. Typical examples are details of background subtraction, normalization schemes, data rejection, region of interest selection, data pre-selection, fitting results, etc. - all of them are constantly changing and so diverse that it appears almost impossible to request them from the scientists in a standardized way. This in turn makes the realization of the FAIR principles for complex non-standardized experiments difficult because FAIR requires a set of metadata which allows to reproduce results from raw data. Key aspects of this metadata within the framework of the FAIR principles have been highlighted by [3] and can by summarized as follows:

- *Adaptability:* Metadata should be flexible to cater to varied researcher needs, allowing customization of data views to highlight specific research elements.
- *Detectability:* Data should be easily accessible, necessitating organized metadata in a structured catalog. Efficient indexing and detailed annotations assist researchers in identifying relevant datasets.
- *Proactivity:* Given the large data output from scientific projects, metadata should effectively manage this volume, ensuring data is accessible and user centric.

## II. SYNCHROTRON USER COMMUNITY

Synchrotron radiation is used as a research tool to investigate structural and dynamic properties of matter in various disciplines, such as materials science, chemistry, biology, engineering, medicine, and environmental studies. The experiments are typically conducted by a user team during a few days of granted beamtime with the users bringing experimental equipment and samples to a specific beamline (experimental station) suited for their scientific question. Some of these beamlines and experimental techniques are highly standardized allowing for a relatively easy capture of metadata but most of the stations and experimental techniques are not easy to standardize. This constitutes a challenge for implementing the FAIR principles [1]. In Germany, the National Research Data

A. Tosson is with the Physics Department, University of Siegen, Walter-Flex-Straße 3, 57072 Siegen, Germany (corresponding author, phone: +49 271 740 3766; fax: +49 271 740 3886; e-mail: amir.tosson@uni-siegen.de).

Mohammad Reza is with the Physics Department, University of Siegen, Walter-Flex-Straße 3, 57072 Siegen, Germany (e-mail: mohammad.reza@uni-siegen.de).

Christian Gutt is with the Physics Department, University of Siegen, Walter-Flex-Straße 3, 57072 Siegen, Germany (e-mail: Christian.gutt@uni-siegen.de).

Infrastructure (NFDI) initiative has been funded with the aim of tackling these problems which are typical in experimental sciences and make data FAIR across the disciplines. The synchrotron user community together with the large-scale synchrotron and neutron facilities is organized in the DAPHNE4NFDI consortium [4].

### III. X-RAY PHOTON CORRELATION SPECTROSCOPY AS A USE CASE

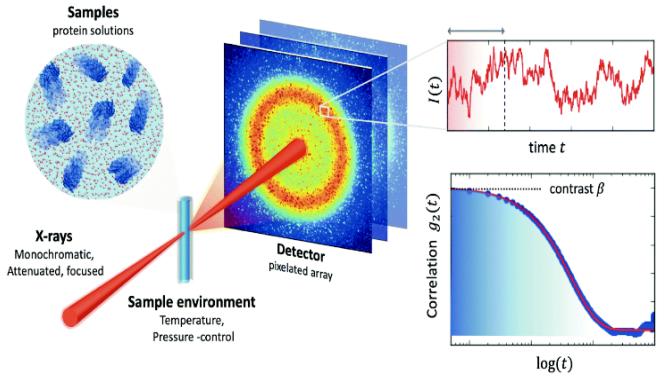

Fig. 1 A typical XPCS experimental setup [5] highlighting the complexity of the experimental and analysis metadata

As a typical use case demonstrating the challenge of metadata collection, we use here the technique of X-ray Photon Correlation Spectroscopy (XPCS) [5]. XPCS is a spectroscopic scattering technique that utilizes the coherence of X-ray beams for studying the dynamics of materials at the nanoscale. By analyzing the temporal fluctuations in the scattered intensity, XPCS can extract information about the motion, relaxation, and structural changes of materials over time. The method thus provides insights into a range of phenomena, from the diffusion of proteins and nanoparticles to the dynamics of polymers and complex fluids, making it a powerful tool for understanding the behavior of materials under various conditions [5]-[7].

A typical experimental setup for XPCS consists of fixed beamline specific components such as X-ray optics and detectors yielding a basic fundamental metadata set which is automatically recorded during the experiment. In contrast, the user specific components consist for example of a sample environment which is engineered to provide control over external parameters, such as temperature or pressure constituting a first set of user specific metadata more difficult to capture. During the experiment, the X-rays are scattered from the sample and the resulting diffraction patterns are captured using a pixelated 2D array detector (see Fig. 1). In this way, a series of patterns are recorded over time which exhibit fluctuations in their intensity mirroring the sample dynamics. Correlating intensities, for example via classical intensity autocorrelation functions in time, yields then information about the temporal evolution of the density-density correlation functions in the sample.

*A. Data Structure and Analysis Pipeline*

The generic XPCS raw data structure consists of time-series of images representing photon intensity over time, where each data point corresponds to a specific intensity, pixel number and time stamp. In a first step, these raw data are preprocessed by removing noise, subtracting background signals, selecting areas of interest, and normalizing the dataset. The central part of the XPCS data analysis is then to calculate the correlation functions (known as g2), which measures how intensity changes over different time intervals for a selected area representing a specific area q in reciprocal space (reciprocal space, i.e., length scale, L ∼ 2 π/q). Upon completing a successful analysis pipeline, one processes the series of images into a 1D correlation function, illustrated in the bottom right of Fig. 1. This curve is generally saved as a 1D array and is often represented by an exponentially decaying function:

$$g_2(q,t) = A + \beta \cdot e^{-2(t/\tau(q))^{\gamma(q)}}$$

where, $\tau$ is the characteristic relaxation time, $\beta$ is the scattering contrast, and $\gamma$ is the stretching exponent called Kohlrausch-Williams-Watts (KWW) exponent. $\gamma = 1$ stands for simple diffusive dynamics, and $\gamma > 1$ indicates a compressed exponential decay, and $\gamma < 1$ stands for the slower dynamics than diffusion.

In another scenario, for a non-equilibrium system, the intensity correlation function (ICF) can be derived as a function of time without temporal averaging. This is expressed in the form of a two-time correlation function (TTC), as given by:

$$C(q, t1, t2) = \frac{<I(q,t1)I(q,t2)>}{<I(q,t1)><I(q,t2)>} - 1$$

where the function C quantifies the progression of the ICF with respect to the average time, t_age, which is calculated as t_age = (t1+t2)/2. TTC is commonly saved as a 2D array. A typical TTC can be seen in Fig. 2.

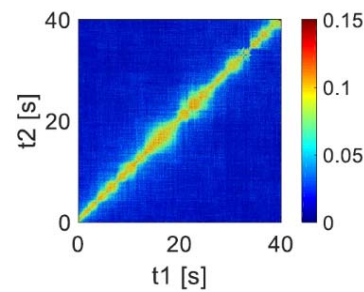

Fig. 2 A typical example of a TTC [8] in which the degree of correlation between images taken at times t1 and t2 is shown as a heatmap

These TTCs are used for further analysis by extracting cuts along specific directions representing correlations as a function of external parameters such as waiting time after externally driving the systems or as a function of X-ray dose, X-ray dose rate or q-value - to name just a few. Obviously, both the production and the further data analysis of such TTCs functions are highly user and experiment dependent which cannot be described by a fixed set of metadata. Other synchrotron and neutron use cases will face a similar problem.

The correlation functions are extensively analyzed both quantitatively and comparatively. They are then fitted to designated models to obtain parameters, notably relaxation times and diffusion coefficients. Fig. 3 presents a high-level overview of the primary elements of the analysis process. Each element encompasses a series of steps and analysis procedures, which vary based on the user and the objective of the experiment and are not standardized. Prior to analysis, the acquired raw data must be preprocessed, which includes improving the signal-to-noise ratio, producing the integrated intensity image, and implementing data masking to identify the regions of interest. After preprocessing, the TTC is generated as previously mentioned. The TTC is then analyzed, its results visualized, and, when relevant, compared or modeled against reference or simulation data. Each step is purposeful and tailored to its application, suggesting the absence of a single standardized analysis approach. Notably, each procedural step provides metadata that describes the performed actions which is crucial for an in-depth understanding and further data reuse.

In practice, the correlation functions fall into one of four categories:
1) Published Data: This refers to data that have already been presented in a scholarly article or publication.
2) Reduced Data (Pre-published): This indicates data that have been processed and is in the final stages of preparation for publication.
3) Simulated Data: These are correlation functions generated through computational modeling or simulation techniques, also potentially training data for machine learning.
4) Reference Data (Calibration Data): This represents theoretical or standard data used for comparison purposes to ensure the accuracy and consistency of measurements.

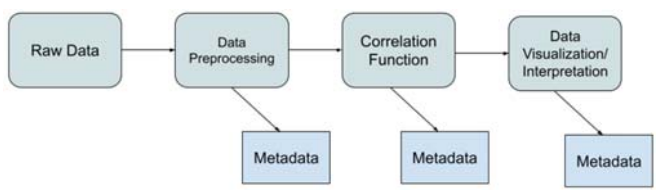

Fig. 3 A typical data analysis pipeline

Beyond the metadata generated within the analytical framework, every correlation function dataset should inherently include essential metadata such as beamline, photon energy, and detector specifics. Furthermore, distinct metadata is requisite based on the category of the correlation function. For instance, published data necessitate an associated article or publication link, while simulated data demand disclosure of the parameters employed in the computational model. The measurement's objective or its applied purpose remain paramount. Certain measurements mandate supplementary details, including dose, dose-rate, specimen age, decay rates, and noise considerations, ensuring meticulous analysis and dependable outcomes.

## IV. State of Art in Metadata

Researchers often emphasize the importance of accurate and complete metadata in various scientific disciplines. A study [9] shows a comprehensive exploration of how people retrieve information from scientific data. It aims to understand the complex concepts of metadata and relevance criteria. The study demonstrates that users considered 45 different types of metadata and used nine relevance criteria to judge the importance of scientific data, providing valuable insights into their information-seeking behavior.

Another study introduces an innovative project aimed at improving metadata quality assessment in the context of Research Data Support, a third-party curation service for researchers [10]. The primary goal was to establish a fair evaluation process for metadata quality. The paper outlines the methodology, which involved single-blind user testing, and presents the results obtained from this experimentation. It also briefly highlights the development and implementation of curation services that followed initial testing.

Open Archives Initiative Protocol for Metadata Harvesting (OAI-PMH) is another approach to harvest metadata [11]. OAI-PMH serves as an easily accessible approach to facilitate repository interoperability, with Data Providers being repositories that make their structured metadata available through this protocol.

In our study, we explore the innovative concept of dynamic metadata, designed to adaptively alter metadata definitions based on evolving needs for documenting data and experimental findings. This method aspires to embody the FAIR data principles for metadata, offering the scientific community a versatile strategy for metadata management.

## V. The Dynamic Metadata Paradigm

### A. An Overview

The Dynamic Metadata Paradigm represents an approach that responds to the key considerations in metadata. It views metadata as a dynamic and continuously developing component, evolving in tandem with the data it represents; capable of adapting alongside the data it describes. In contrast to the traditional metadata which functions as a fixed descriptor for datasets, this paradigm offers a more flexible and adaptive role for metadata. To enhance the tailorability, Dynamic Metadata leverages the concept of the Internet of Things (IoT) to enable community members to easily share their own metadata items and content, such as descriptions, tags, or annotations, within a unified framework or a cloud-based schema. This central metadata repository serves as a collective pool of knowledge and resources, offering a more collaborative and efficient environment for the community.

A vital component of dynamic metadata lies in its foundation on cutting-edge indexed database technologies. By efficiently organizing these databases adeptly, one can substantially enhance the metadata's detectability. To achieve a balance between adaptability and detectability, it is recommended to employ a mix of relational and non-relational database technologies. This concept will be explored in the following sections.

To ensure proactive handling of metadata under high-rate data scenarios, dynamic metadata recommends a framework architecture where data and metadata are maintained on two separated systems (i.e., MetaData Manager (MDM) and primary data manager) that can seamlessly interact with each other, as explained in the following section.

### B. The Proposed Dual-System Framework

Considering the complexities of managing both data and its associated metadata, dynamic metadata suggests a dual-system framework:

- *MetaData Manager (MDM)*
- *Role*: Exclusively dedicated to managing metadata, it oversees the collection, storage, update, and retrieval of all metadata.
- *Advantages*: By focusing solely on metadata, the MDM can employ specialized algorithms and storage structures optimized for metadata's unique characteristics. This can lead to faster queries, updates, and better organization.
- *Primary Data Manager:*
- *Role*: It is directly responsible for handling the actual data, ensuring its integrity, storage, retrieval, and backup.
- *Advantages*: Without the overhead of managing metadata, the primary data manager can provide faster and more efficient data operations.

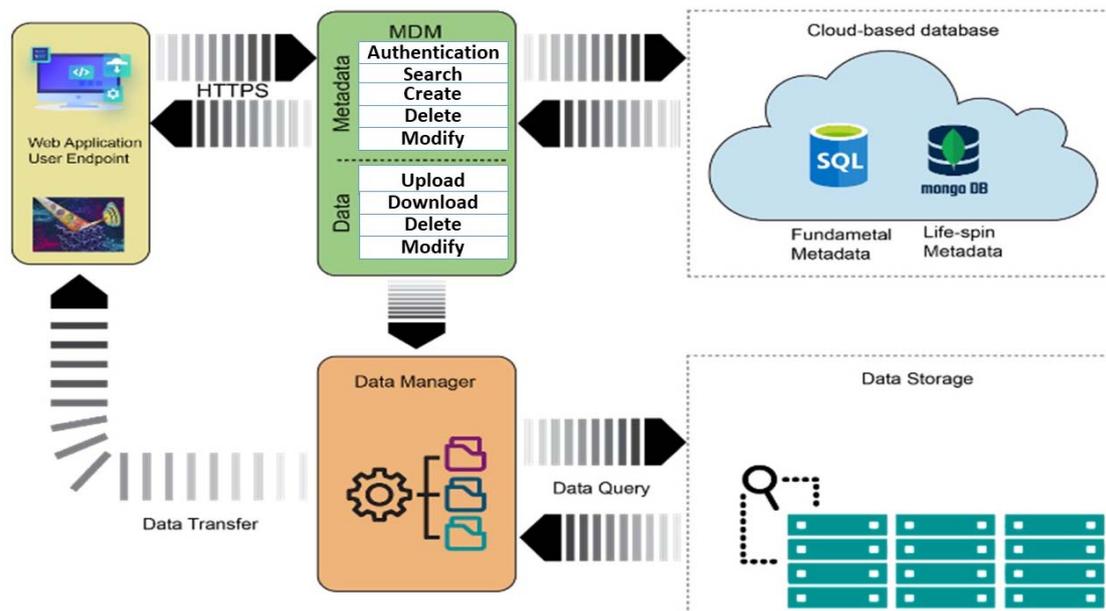

Fig. 4 A scheme of the proposed dual-system

Fig. 4 illustrates the schematic of the proposed system, comprising two interlinked subsystems. It is composed of a web application user interface that connects via HTTPS. The core module, denoted as "MDM," manages both metadata and data. For metadata operations, the MDM offers features like authentication, search, creation, deletion, and editing. It connects to a cloud database containing two metadata categories: "Fundamental Metadata" and "Life-spin Metadata". For data file operations, functions like upload, download, edit, and delete are supported. These files are directed to the Data Manager, overseeing data transfer and file queries. This manager then communicates with the specified data storage, referred to as the data lake.

### C. The Anatomy of the Dynamic Metadata

This section demonstrates the principles of the dynamic metadata approach and elaborates on how the MDM structures and oversees its management. The core of this schema is its structural composition, which consists of two main components: the fundamental metadata and the life-spin metadata.

Fundamental Metadata

The fundamental metadata serves as the foundational support for the platform's operations. Provided directly by the platform, it adheres to specific rules within a designated namespace, which facilitates communication between the MDM and the primary data manager. It is systematically structured in a tabulated manner, ensuring efficient and consistent data storage and retrieval. Essentially, it directs the platform's data management and presentation, ensuring seamless and reliable operations.

In technical terms, fundamental metadata falls under the category of administrative metadata. It provides details such as data file properties, data location, access control, ownership and rights, retention, and disposal, and more. These metadata items are required whenever a corresponding task is carried out. That is why it is recommended to keep them as minimal as possible. This helps lighten the load for users and makes for a better user experience.

Life-Spin Metadata

The life-span metadata functions as a collaborative hub,

enabling users to collectively contribute to and refine a continually evolving metadata schema. While the fundamental metadata adheres to a rigid set of predefined rules and structures, the life-span metadata offers a more participative approach. It encourages users to input metadata items based on their unique perspectives and experiences. Moreover, this communal space fosters a feedback-rich environment, where members can share insights and suggestions, facilitating a collective enhancement of the metadata schema.

The MDM operates as a version control system, allowing users to track, oversee, and access different versions of the life-span metadata. In addition to supporting Create, Read, Update, and Delete (CRUD) actions for items they own, it also facilitates commenting and ranking on items contributed by other users.

In technical terms, Life-Spin metadata covers two key categories: descriptive and structural Metadata. Descriptive Metadata provides comprehensive information about a digital resource, such as titles, authors, dates, keywords, and other elements. Meanwhile, Structural Metadata dives into the organization and connections among digital resources, specifying how various components are structured and linked.

To simplify and make Schema documentation more flexible, it is recommended to use JSON data representation. JSON is not tied to any specific programming language and is a commonly used format in many applications' API outputs. It employs key-value pairs to create a map-like structure. The key is a string that identifies the pair, while the value contains the information associated with that key. Furthermore, JSON parsing offers a safer approach compared to other data representations [12].

### D. Definitions and Concepts

Concept Hierarchy

Dynamic metadata primarily depends on two main categories: the method utilized and the object that is requested to be created.

*The method* involves experimental approaches in different settings like radiation facilities or home laboratories (e.g., XPCS, XSAF, and SAXS).

*The object* represents the created entity (such as a data file, attachment, graphics, etc.), which results from a user's creation action.

Within the hierarchy, the method holds a higher position than the process. For each method, there should be a specific set of metadata schemas that match the objects that can be created through the platform.

Metadata Concepts

*Item:* Typically, it refers to a single unit or piece of metadata. Items are individual elements of metadata that provide descriptive information about a particular object or resource. These items can include details such as titles, authors, dates, keywords, and other attributes.

*Instance:* An instance, in a conceptual sense, acts as a container responsible for storing or dynamically generating metadata items related to a specific method, task, or object.

*Unique Identifier:* A unique identifier is a special code, name, or value used to clearly identify a specific instance. This identifier is designed to ensure that there is no ambiguity or confusion when identifying and referencing that specific instance.

### E. The Naming System

When it comes to naming elements in JSON, it is a widely accepted best practice to adopt a uniform naming convention. This helps guarantee clarity and ease of maintenance.

In the context of this research, we strongly recommend and consistently employ the snake case naming convention. Snake case entails a specific naming pattern where spaces between words are replaced with underscore (_) characters, and the initial letter of each word is written in lowercase; for example: "first_name", "last_name", "phone_number".

### F. The Technical Description

Fig. 5 provides a detailed perspective on the role of metadata within the community. This community consists of researchers who utilize "Method A" for their studies. Within Method A, three distinct objects, namely X, Y, and Z, can be created, each leading to its own unique dynamic metadata schema within the system. The dynamic metadata for each Method and object is defined by a specific UI.

*Creation Requests:* These are requests forwarded by users to the MDM. A valid request must include a UI that represents both the method (like Method A) and the object (such as X, Y, or Z) they wish to create. Such requests are then redirected to their respective tasks within Method A. Administrative metadata (i.e., fundamental metadata) related to the object must be included as well. If the request meets the necessary requirements, it will proceed to the endpoint, leading to the creation of the required object. Subsequently, an instance of the object's current dynamic metadata will be initialized, providing access to the life-span metadata repository.

*Fundamental and Life-Spin Metadata:* The 'Fundamental Metadata' is the essential data elements which are extracted directly from the user's request. These elements are static and unchanging, thereby ensuring a uniform and consistent framework for all stakeholders. For security and ease of reference, these elements will be stored and backed up in a tabular format. On the other hand, the 'Life-Spin Metadata' is housed in a cloud infrastructure. Once access is granted, users have the flexibility to make alterations to this life-span metadata. This allows them to represent their findings, insights, and experiences within a particular method and creation process.

*On Collaboration and Community Sharing:* The innovation of this system lies in its community-centric design. Within the broader scope of the Method and Object, users are granted the flexibility to share, edit, or remove their life-span metadata items. To promote the collaborative environment, users can comment on, and rank life-spin metadata items created by their peers.

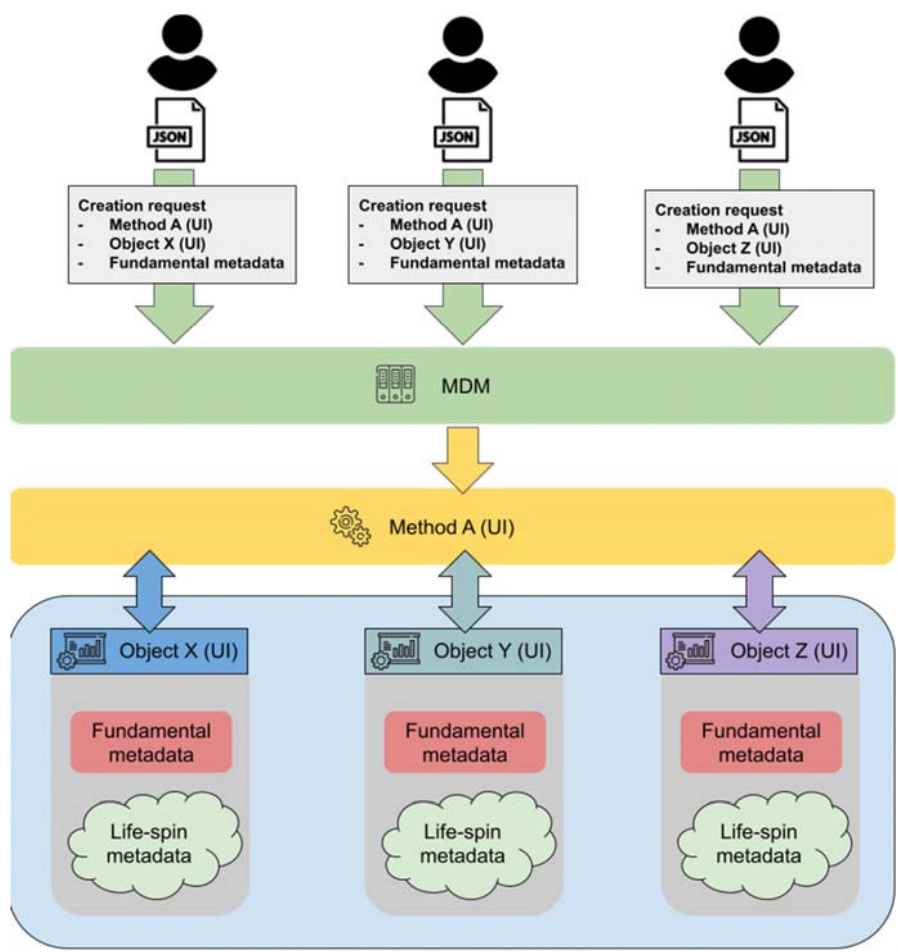

Fig. 5 The Flow of Metadata Interaction within the Method A Community - Users' Requests (X, Y, Z), Central Metadata Management, and Collaborative Life-Spin Metadata Updates

## VI. Application of Dynamic Metadata in Use Case XPCS

In the context of our use case XPCS we assume that four object types need to be created. 1) The experiment instance serves as a category to collect dataset instances such as the Electronic Lab Notebook (ELN) or sample details related to a specific measurement or the entire beamtime. 2) The dataset instance acts as a repository for data files (like correlation function data) and can also include further information, attachments, and code snippets. 3) The ELN serves as a hub for both user and facility-provided metadata and information. 4) The sample instance describes the sample used in the experiment ideally linked to further sample identifiers. The MDM is expected to feature four distinct dynamic metadata schemes, each corresponding to the four objects within the XPCS framework. However, it is important to note that the number of objects and their respective schemes might vary depending on the method employed and the specific requirements of the community.

For a better organization and higher efficiency, we suggest implementing a system hierarchy tailored for XPCS, anchored on dynamic metadata, as shown in Fig. 6. In this system, *XPCS users* are on the top of the hierarchy representing the primary users or stakeholders of this system. They can create multiple "*experiment*" objects. Each object links to a particular beamline or a specific experiment performed at a synchrotron facility. Fundamentally, each experiment object is characterized by its name, date, facility, and owner. Upon initialization, users gain access to the community metadata repository, also known as the life-span metadata. They have the choice to either utilize its current version or update it to suit their specific needs. Importantly, any modifications or actions are shared across the entire XPCS user community, fostering a collaborative environment. Within the same experiment objects users can initialize three objects:

- The "*dataset*" object contains the details about the actual datasets. Users can create multiple dataset objects within the same experiment, depending on the specific details of the experiment and their personal preferences. For example, they might have a separate dataset for each day of the experiment or for each sample measured. Inside these dataset objects, users can include data files (in our case, correlation functions data), attached files (e.g. graphs or reports), and code snippets to show others how to process this data. At its core, each dataset object is defined by its name, the experiment it is connected to, and its creation date. Using the same method, after setting things up, users can tap into the shared repository of life-span metadata specific to the dataset object.
- The "*sample*" objects serve as categories to gather data about the samples measured. Within a single experiment, users can create numerous sample objects. Each sample object is characterized by its name, creation date, and the associated parent experiment. Additional details about the sample, such as its manufacturer, composition, and production, are allocated to the life-span metadata. In standard practice, samples are usually associated with a Persistent Identifier (PID), provided by entities like the DAPHNE4NFDi consortium.
- The "*ELN*" object serves as a repository for collecting, filtering, and organizing experimental metadata. This metadata can be input manually by users or received automatically from the facility. The ELN maintains both Fundamental and Life-Spin Metadata, ensuring comprehensive documentation of experimental procedures, modifications, and updates. Technically, in addition to allowing users to input details and information, the ELN should offer features for filtering and organizing the gathered metadata and present them in a manner easily comprehensible to humans.

The developed XPCS platform is based on a dual-system approach and dynamic metadata integration. This platform is a web-based application using a microservice architecture. Fig. 7 displays the platform's Unified Modeling Language (UML) representation, emphasizing its design tailored for managing XPCS data, specifically the correlation function datasets. At the user interface layer, interactions start with HTTP requests. These can be separated into two main channels: Data File operations and Metadata management. The former incorporates a Data File Manager with capabilities to save, verify, load, and delete data files. These data files reside in a specialized Data Lake, optimized for correlation function files. Concurrently, the Metadata section employs an MDM system, featuring four core classes: Experiment, Dataset, Sample, and ELN. Each class possesses distinct attributes such as name, date, type, and owner ID and accommodates CRUD operations, augmented with commenting and ranking functionalities. The metadata is systematically stored across two databases: the Fundamental Metadata Database and the Life-Spin Metadata Database.

User queries are bifurcated, and metadata is directed to the MDM, while data files proceed to the file manager. These files undergo validation and verification according to community-agreed criteria, focusing on size, structure, and type. Post-validation, the files are stored in a data lake. Subsequently, a unique identifier (UID) is generated, which is relayed to the MDM and associated with its respective metadata.

Queries pertaining to metadata are directed to the MDM. They are processed according to their classification (Fundamental or Life-spine), the associated entity (experiment, dataset, sample, or ELN), and the specified action (CRUD operation, commenting, or ranking). The MDM oversees databases for both metadata types. The life-spin metadata database serves as a communal repository among users, supervised by the MDM. Following each user query, the database is refreshed, allowing users to engage with the updated version. Additionally, the MDM offers a search functionality for effortless metadata item retrieval. Furthermore, a version control mechanism is in development to monitor the various versions over time.

## VII. ADDED VALUES

The flexibility of dynamic metadata is a significant advantage for realizing FAIR data in experimental sciences with highly dynamic variable setups, samples, and analysis schemes. In such instances traditional fixed metadata approaches may have difficulties to keep up with rapidly changing data characteristics while dynamic metadata can easily be updated and expanded as new insights emerge or data formats evolve. In experiments that involve real-time data collection, dynamic metadata systems can adapt descriptions and annotations in real-time, ensuring that researchers and analysts always have access to the most relevant information. Furthermore, dynamic metadata contributes to improved data quality by providing contextual information about data sources, transformation processes, and lineage. This transparency enhances trust in the data and supports effective data governance. Dynamic metadata also fosters enhanced collaboration among researchers, as it allows for easy sharing and integration of data while adapting to new contexts and interpretations. This can promote a new concept known as "FAIR metadata.".

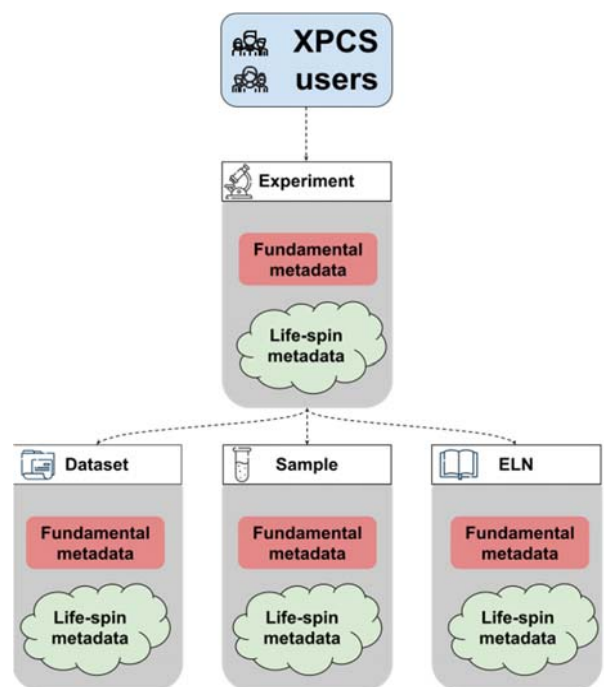

Fig. 6 Scheme of the concept hierarchy

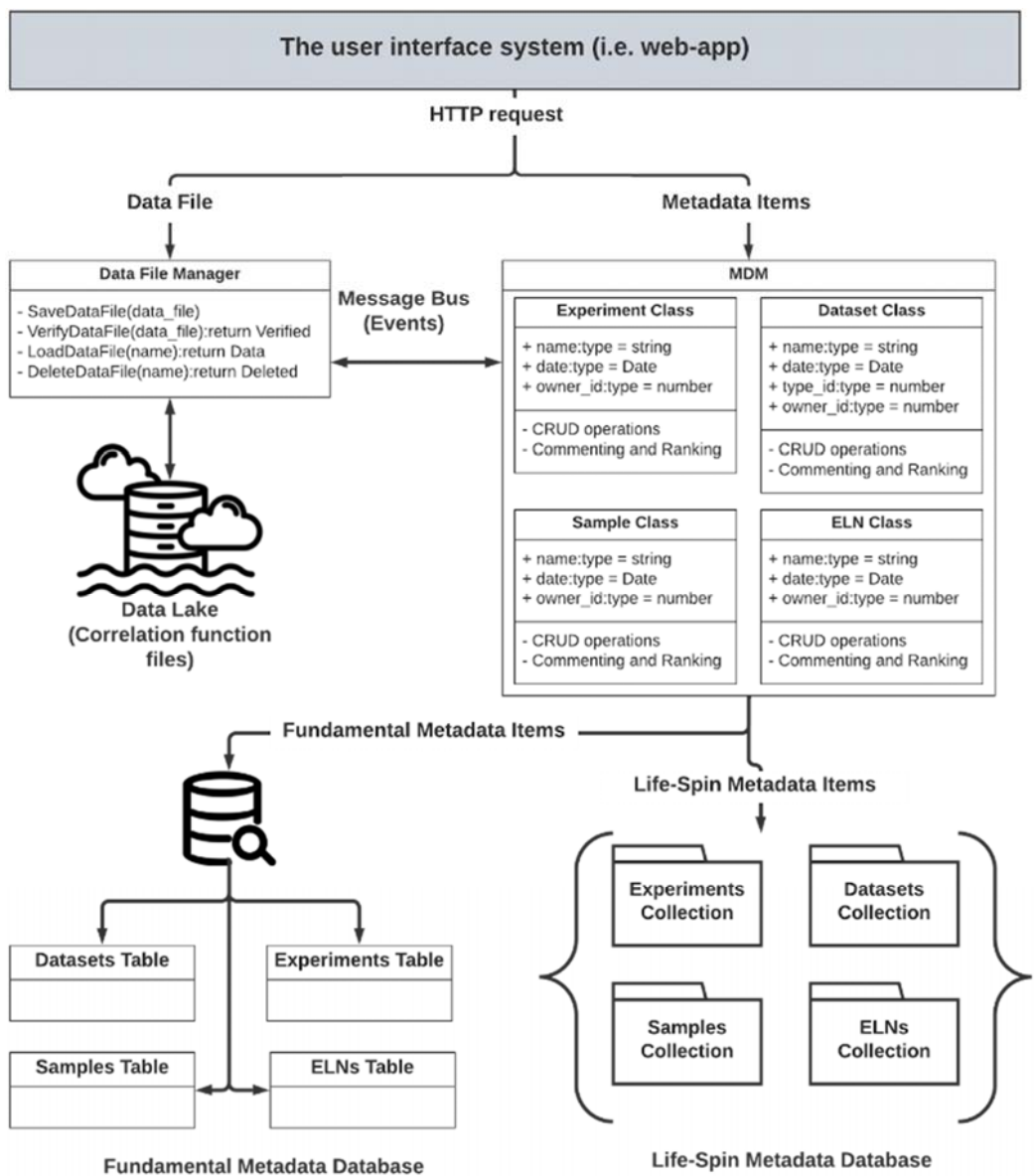

Fig. 7 UML design for the XPCS platform

## VIII. CONCLUSION

Metadata plays a crucial role in enhancing data management across many research fields, particularly in synchrotron experiments spanning various scientific disciplines. Given the continuously changing nature of such experiments, static metadata definitions may fall short for realizing the FAIR principle. In this article, we proposed the concept of dynamic metadata, using XPCS as a representative use case. This approach allows researchers to adapt metadata definitions according to the evolving needs of diverse experiments. Implementing dynamic metadata has the potential to enhance data acquisition, sharing, analysis workflows, reproducibility, interoperability, and knowledge dissemination. We presented a dual-system to address metadata complexity and provided an architectural outline of the developing XPCS platform, exemplifying the dynamic metadata approach and the dual-system.

## ACKNOWLEDGMENT

We would like to express our profound gratitude to the Deutsche Forschungsgemeinschaft (DFG) and the Bundesministerium für Bildung und Forschung (BMBF) for their generous financial support that made this research possible.

Special thanks to our esteemed partners in DAPHNE4NFDI, whose continued collaboration and insights were instrumental in driving the progress and outcomes of this study. We are equally grateful to the technical team at DESY for their invaluable assistance and expertise throughout the course of our research.